\documentclass[11pt,letterpaper]{article}
\usepackage[T1]{fontenc}
\usepackage[utf8]{inputenc}
\usepackage{lmodern}
\usepackage{upgreek}
\usepackage{textcomp}
\usepackage{amsmath,amsfonts,amssymb,bm}
\usepackage{amsmath,amsfonts,amssymb,bm}
\usepackage{amsthm}
\usepackage{mathabx}
\usepackage{mathrsfs}
\usepackage[english]{babel}
\usepackage{indentfirst,setspace,fancyhdr}
\usepackage[outercaption]{sidecap}
\usepackage[footnotesize]{caption}
\usepackage{geometry}
\usepackage{epstopdf}
\usepackage{color}
\usepackage[none]{hyphenat}
\usepackage[toc,page]{appendix}
\usepackage{multicol}
\usepackage{colortbl}
\usepackage{multirow}
\usepackage{array}
\usepackage{pstricks}
\usepackage{wrapfig}
\usepackage{subcaption}
\usepackage{fancyref}
\usepackage{pdfpages}
\usepackage{mathtools}
\usepackage{booktabs} 
\usepackage{cite}
\usepackage{quoting}
\usepackage{longtable}
\usepackage{cancel}
\usepackage{makecell}
\usepackage{tikz,stackengine}
\usetikzlibrary{matrix,calc}
\usepackage{mathtools}
\usepackage{blkarray}
\usetikzlibrary{decorations.pathreplacing}

\usepackage{pdflscape}
\usepackage{chngcntr}

\newtheorem{proposition}{Proposition}

\definecolor{Gray}{rgb}{0.5,0.5,0.5}
\definecolor{darkblue}{rgb}{0.12,0.15,0.49}
\definecolor{light-gray}{gray}{0.85}
\definecolor{mygreen}{rgb}{0, 0.5, 0}
\makeatletter
\renewcommand\@biblabel[1]{#1.} 
\makeatother
\allowdisplaybreaks

\newcommand{\xddots}{%
	\raise 4pt \hbox {.}
	\mkern 6mu
	\raise 1pt \hbox {.}
	\mkern 6mu
	\raise -2pt \hbox {.}
}
\emergencystretch=\maxdimen
\DeclareFontFamily{OT1}{pzc}{}
\DeclareFontShape{OT1}{pzc}{m}{it}{<-> s * [1.10] pzcmi7t}{}
\DeclareMathAlphabet{\mathpzc}{OT1}{pzc}{m}{it}

\graphicspath{{Figures/}} 
\newcolumntype{g}{>{\columncolor{light-gray}}c}

\makeatletter
\renewcommand{\maketitle} 
{ \begingroup \vskip 10pt \begin{center} \large {\bf \@title}
	\vskip 10pt \large \@author \hskip 20pt \@date \end{center}
  \vskip 10pt \endgroup \setcounter{footnote}{0} }
\makeatother 
 
\let\baraccent=\= 
\renewcommand{\=}[1]{\stackrel{#1}{=}} 

\usepackage{ulem}
\newcommand\dsout{\bgroup\markoverwith{\textcolor{red}{\rule[0.5ex]{1pt}{1pt}}}\ULon} 

\begin{document}

\title{The pursuit of happiness}

\author{Debora Princepe$^{1,2}$, Onofrio Mazzarisi$^{1,2,3}$, Erol Ak\c{c}ay$^{3}$, \\
Simon A. Levin$^{5,6}$ and Matteo Marsili$^{1,2}$\\
~~\\
{\small $^1$The Abdus Salam International Centre for Theoretical Physics (ICTP), Trieste 34151, Italy.}\\
{\small $^2$The Laboratory for Quantitative Sustainability, Viale Miramare 24/4, Trieste 34135, Italy}\\
{\small $^3$National Institute of Oceanography and Applied Geophysics (OGS), Trieste 34151, Italy}\\
{\small $^4$Department of Biology, University of Pennsylvania
433 S. University Ave Philadelphia, PA 19104}\\
{\small $^5$Department of Ecology and Evolutionary Biology, Princeton University, 
NJ 08544, USA}\\
{\small $^6$ High Meadows Environmental Institute, Princeton University, NJ 08544, USA}\\
~~\\
~~}



\maketitle


\begin{abstract}
This paper studies populations of individuals -- that we call {\it homo-felix} -- who maximise an objective function that we call {\it happiness}. The happiness of one individual depends on the payoffs that they receive in games they play with their peers as well as on the happiness of the peers they interact with, and on their status. Individuals care more or less about others depending on whether that makes them more or less happy. This paper analyses the ``happiness feedback loops'' that result from these interactions in simple settings. We find that individuals revise their other-regarding attitudes in terms of relative comparisons. In particular, 
individuals tend to care more about individuals who are happier than them, while spiteful behavior may emerge only under poor socio-economic conditions and is directed towards the worst off individuals. 
Such general results are illustrated by studying simple $2\times 2$ games, where we find that {\it homo-felix} can converge to a  variety of equilibria which includes but goes beyond Nash equilibria, and in {an $n$-persons public good game}.
\end{abstract}

\begin{quote}
    How selfish soever man may be supposed, there are evidently some principles in his nature, which interest him in the fortune of others, and render their happiness necessary to him, though he derives nothing from it, except the pleasure of seeing it.
    
    \rightline{[Adam Smith, The Theory of Moral Sentiments, 1759]}
\end{quote}

Adam Smith is most known for unveiling the working of the {\it invisible hand} whereby individuals pursuing their own self-interest also promote the welfare of others through markets, thereby contributing to the common good. This mechanism, as he pointed out, is embedded within a social fabric shaped by moral sentiments. Yet moral sentiments and  concerns for others have no place in economic theories~\cite{mas1995microeconomic} that formalised his invisible hand. This is 
the very framework that has shaped our institutions and the modern world. Within this paradigm, individuals are modeled as rational agents driven by self-interest, responding only to rewards and punishments~\cite{fudenberg1991game,mas1995microeconomic}. As this framework has come to dominate, the moral foundations that once underpinned economic life have been increasingly 'crowded out'~\cite{bowles2016moral}.

A wealth of empirical evidence~\cite{fehr2003nature,fehr2006economics} demonstrates that human behavior consistently deviate from payoff-maximizing behavior, being profoundly influenced by social considerations~\cite{helliwell2017social} (e.g. fairness and reciprocity) and the well-being of others~\cite{helliwell2025caring}. Such evidence spans from anthropological studies\cite{mauss1925ledon,bowles2016moral}, economics~\cite{putnam1993,fehr2006economics,feigenberg2013economic}, social sciences~\cite{ostrom1990governing,finlayson2005invisible,diener1999subjective,fowler2008dynamic} and is supported by findings in neuroscience~\cite{izuma2018,marsh2022} and evolutionary arguments~\cite{nowak2006,bowles2011cooperative}, as discussed in Section~\ref{sec:evidence}. 
These findings have motivated several theoretical approaches 
in the attempt to explain them~\cite{fehr2006economics,tilman2019localized,crabtree2024influential,dixit2017social} or to show how the concern for others may emerge endogenously in interacting populations~\cite{akccay2009theory,alger2019,alger2023evolutionarily}. 

This paper presents a theoretical framework for the emergence of other-regarding preferences which radically departs from these approaches. Indeed, to the best of our knowledge, all these attempts consider  individuals that optimize a combination of their own and their opponents payoffs. Here we discuss 
a perspective on individual behavior -- which we call {\it homo-felix} -- that assumes that individuals maximise an objective function -- that we shall call happiness -- that is a combination of their own payoffs and the happiness of others. 
The consideration for others is an adaptive trait, in the sense that individuals care for the happiness of others, either positively or negatively, to the extent that it makes them happier. 

The use of the term ``happiness'' rather than ``utility'' reflects a {\it eudaimonic} conception of well-being~\cite{bruni2005economics,ryan2001happiness}, which departs from the hedonic view by emphasizing its social dimension. In simple terms, our framework aims at capturing the idea that a component of what makes someone ``happy'' is that the people he/she interacts with are also ``happy'', as suggested by Adam Smith in the sentence above. One key feature of what we call ``happiness'' is that it should be observable. In this sense, happiness may correlate with a measure of social status or reputation. 
The fact that in this framework individual happiness depends on the happiness of others, to the extent that that makes them happier, introduces non-trivial feedback mechanisms that renders happiness an emergent collective property with a fundamentally social dimension. This feedback and its consequences is the main object of study in this paper. 

Within this framework, we find that a concern for reciprocity emerges endogenously and that individuals adjust their behavior according to measures of relative well-being. 
Most well-off individuals fail to develop pro-social preferences, while worse off individuals do. Spiteful behavior does not emerge when the average well-being is high and, when it is low, it is directed towards the worst off individuals. 

After defining our set-up, we first articulate these findings by drawing some general conclusions, which are then illustrated by a number of prototypical examples of two-players and $n$-players games. Along with the usual Nash equilibria, we shall find other equilibria, that qualitatively agree with findings in experimental economics~\cite{fehr2006economics}. 

We conclude with a general discussion.

\section{The evidence on pro-social behavior}
\label{sec:evidence}
Mauss's seminal work~\cite{mauss1925ledon} reveals that in many archaic societies economic transactions occur in the form of gifts, where gift-giving is a complex social practice that fosters bonds and mutual obligations. Similarly, Putnam observed~\cite{putnam1993} that networks of relationships and trust within communities -- the so-called {\it social capital} -- significantly impact public affairs and collective well-being, suggesting that individuals' actions are deeply embedded in social contexts, where considerations of others play a crucial role\footnote{The use of the term "capital" in terms like social or human capital is itself indicative of the extent to which the {\it homo-economicus} narrative has permeated economic discourse. As if social relations or education do not have a value in themselves but they are acquired as means to achieve economic benefits.}.
Elinor Ostrom showed that communities can sustain cooperation and manage shared resources  through decentralized mechanisms grounded in other-regarding attitudes~\cite{ostrom1990governing}, an insight that has been formalized and extended through game theoretical approaches~\cite{tilman2019localized,crabtree2024influential}. 
These informal governance structures are thought to have laid the groundwork for more formal legal systems -- such as the merchant law of medieval Europe~\cite{milgrom1990role} -- which emerged to expand cooperation to strangers~\cite{bussani2018strangers}. In a parallel manner, modern microcredit schemes have shown that, within tightly-knit communities, social capital can effectively substitute for conventional financial collateral, fostering a virtuous cycle of trust and economic empowerment~\cite{feigenberg2013economic}.

These evidences underscore that concerns for others are not marginal anomalies but core features of human economic life
. Efforts to reconcile choice behavior that involve consideration for others, with the {\it homo-economicus} approach of neo-classical economics have been very influential~\cite{becker1981treatise}. In this view, individuals are assumed to make life choices -- whom to marry, how many children to have, how much time to invest in relationships and in education -- as outcomes of intertemporal optimization problems, seeking to maximize expected utility subject to constraints on time, income, and information. In this view, individuals engage in relationships as well as in education not for their own sake, but rather as a result of investment choices aimed at acquiring {\it social capital} or {\it human capital} that will increase future expected returns. 
This approach offers elegant models of social behavior with far-reaching insights, and it has become a dominant social narrative, shaping societal and political organization by promoting the idea that individual self-interest leads to optimal social outcomes~\cite{finlayson2005invisible}.

Several authors have criticized this approach, because it requires well-defined and stable preferences, the ability to forecast long-term consequences, and a remarkable level of rational sophistication, all features which are inconsistent with observed human behavior~\cite{simon1982models,kahneman1979prospect}. It has also been pointed out that, if utility is defined broadly enough to include any possible motivation, like altruism, self-sacrifice, or moral duty, then the theory risks becoming tautological, losing explanatory power~\cite{sen1977rational}. Furthermore, J. Stiglitz~\cite{stiglitz2019} observed that ``Rational actors, pursuing their self-interest, have produced an irrational, unsustainable economy'', with soaring levels of inequality~\cite{piketty2014}. 
Yet policies that promote pro-social attitudes, in spite of their ability to lead to better collective outcomes~\cite{ostrom1990governing,betsch2013inviting,graf2024public}, face critiques of naivety within the dominant narrative of {\it homo-economicus}, given their lack of a formal micro-economic foundation.

The {\it homo-economicus} paradigm is grounded in a hedonic view in which well-being is equated with the attainment of pleasure and the avoidance of pain~\cite{ryan2001happiness}, operationalized in the maximization of utility, wealth and/or income. Although income is essential for well-being in conditions of scarcity, because it enables access to basic resources, its correlation with well-being diminishes as income rises~\cite{kahneman2010high}. Empirical research shows that income and wealth provide an incomplete measure of well-being, a fact illustrated by ``happiness paradoxes''~\cite{bruni2005economics}, such as the stagnation of self-reported US happiness despite a $60\%$ rise in real per capita income in the period 1946–1970. A large body of research has focused on the determinants of well-being (see Refs.~\cite{diener1999subjective,ryan2001happiness,bruni2005economics,kahneman2006developments}). Among the relevant components, social connectedness and relatedness~\cite{ryan2001happiness,helliwell2017social} is close to the top of the list. This aligns with an {\it eudaimonic} perspective (see~\cite{ryan2001happiness,bruni2005economics}) in which happiness does not coincide with subjective pleasure, but rather with the expression of virtue, rooted on Aristotelian ethics according to which ``the happy man needs friends'' (Nic. Ethics IX, 9, 1169b). The social dimension of happiness is also prominent in Makiguchi's theory~\cite{bethel1973makiguchi}, that equates happiness with value creation, where value entails 
individual satisfaction -- derived from Beauty (sensory pleasure) and Benefit (existential gain) -- and Good, which consists in creating value for others.

Empirical research increasingly supports the idea that well-being is positively correlated with altruistic behaviour~\cite{helliwell2025caring} and is socially contagious, with an individual's likelihood of being happy increasing significantly if a close friend or relative becomes happier~\cite{fowler2008dynamic}. Similar findings have emerged in digital contexts, where positive and negative affect propagate through online social networks~\cite{kramer2014experimental}. These insights suggest that well-being is not merely an individual attribute but a dynamic, interdependent phenomenon embedded in the social fabric. Likewise, prosocial behavior is strongly influenced by social norms~\cite{bockler2016structure}, and it is  entrenched in culture and religions. For example, spiritual priming can significantly increase generosity~\cite{shariff2007}. 


Besides its social dimension, prosocial behavior is rooted in a genuine altruistic attitude of our neural hardware~\cite{marsh2022}. Indeed research in neuroscience has shown that altruistic behavior builds on the capacity to perceive the emotional status of others~\cite{decety2008emotion,nummenmaa2012emotions}. 
Exposure to others’ emotions can evoke the same emotional
responses in the self~\cite{decety2008emotion}.
Several studies have shown that altruistic actions activate the brain's reward system, suggesting that prosocial behavior is inherently rewarding~\cite{izuma2018,lozada2011beneficial}. Punishment or observing the failure of envied ones also activates the reward system. Yet, while the empathic concern implicit in genuine altruism entails the internalization of the emotional status of the other~\cite{singer2006empathic}, e.g. feeling the other's pain as one's own, the brain sharply distinguishes between self and an objectified other in the case of punishment~\cite{buckholtz2008neural}.

Integrating these insights into game theory is essential for modeling human behavior. Several interesting efforts have been made in this direction. For example, Tilman {\it et al.}~\cite{tilman2019localized} show that spatially localized prosocial behaviors significantly impact the sustainability of shared resources. Crabtree {\it et al.}~\cite{crabtree2024influential} instead focus on how different governance structures affect prosocial behaviors. Alger and Weibull~\cite{alger2019,alger2023evolutionarily} have shown that Kantian moral behavior -- one that the individual would prefer to be universally adopted -- can emerge from evolutionary dynamics. Similarly, Akcay {\it et al.}~\cite{akccay2009theory} show that prosocial behavior can emerge from evolutionary forces, whereby emotions and social norms can act as proximate drivers, ultimately serving fitness interests. The model we discuss in the next Section offers a different framework for the endogenous emergence of other-regarding attitudes.

\section{{\it homo-felix}\,: definitions}

Consider a set $\mathcal{N}$ of $n=|\mathcal{N}|$ individuals. Each of them occupies one of the nodes of a graph $(\mathcal{N},\mathcal{V})$, with $\mathcal{V}\subseteq\mathcal{N}\times\mathcal{N}$ being the set of edges. Let $\partial_i\subset\mathcal{V}$ be the neighborhood of $i$, i.e. the set of individuals $j$ such that $(i,j)\in\mathcal{V}$. Each individual $i$ engages her neighbors $j\in\partial_i$ in a game, which is defined by the set of strategies $\mathcal{S}_i$ she can adopt, and by a payoff matrix
$\pi_i(s_i,s_{-i})$ that depends on her strategy $s_i\in\mathcal{S}_i$ and on the profile $s_{-i}=\{s_j\in\mathcal{S}_j,~\forall j\in\partial_i\}$ of opponents' strategies. 

We consider individuals with possibly other-regarding preferences who choose their strategies $s_i$ in order to maximize an 
objective function -- that we shall call ``happiness'' -- that reads
\begin{equation}
\label{eq:defhf}
    u_i(s_i,s_{-i};\vec p_i)=\left[1-q_i(\vec p_i)\right]\pi_i(s_i,s_{-i})+\sum_{j\in\partial_i} p_{i,j} u_j(s_j,s_{-j};\vec p_j)+h_i\,.
\end{equation}
Here $\vec p_i$ is a vector of coefficients $p_{i,j}$ that sets the degree to which happiness $u_j$ of $j$ contributes to that of $i$, $h_i$ is a baseline happiness, and $q_i$ is a continuous function of $\vec p_i$ that captures the trade-off between material wealth and the happiness of others. Eq.~(\ref{eq:defhf}) captures three components of subjective well being described in empirical research~\cite{helliwell2017social}: material wealth or income, social support, and other factors. The net effect of these latter factors is condensed in the variable $h_i$, which we shall call ``status'' for simplicity, and it may represent social and environmental conditions (e.g. health and education, freedom and autonomy, security and trust) that impact well-being.

The trade-off in Eq.~\eqref{eq:defhf} between individual payoffs and the happiness derived from others is subject to the following conditions:
\begin{eqnarray}
\pi_i(s_i,s_{-i})&\ge& 0\,,\qquad \forall s_i\in\mathcal{S}_i,~~s_{-i}\in\mathcal{S}_{-i}\label{c1}\,,~~\forall i\in\mathcal{N},\\
\vec p_{i}&\in & \left\{\vec p\in\mathbb{R}^{n}:~\sum_{j\in\partial_i}|p_{i,j}|<1;~q_i(\vec p_i)\in [0,1];\,p_{i,j}=0~~\forall j\not\in\partial_i\right\},~~\forall i\in\mathcal{N},
\label{c2}\\
q_i(0) &=& 0\,,\qquad\forall i\in\mathcal{N},\label{c3}\\
p_{i,j}\frac{\partial q_i}{\partial p_{i,j}} &> & 0\,, \qquad \forall i\in\mathcal{N},~~\forall j\in\partial_i\,.\label{c4}
\end{eqnarray}
These conditions ensure {\it i)} that any other-regarding behavior $\vec p_i\not=0$ is costly (Eqs. \ref{c1}, \ref{c4}), i.e. that it reduces the happiness derived from payoffs, {\it ii)} that other-regarding behaviors are bounded and limited to those that keep happiness a non-decreasing function of payoffs (Eq.~\ref{c2}), 
and {\it iii)} that selfish individuals ($\vec p_i=0$) conform to the predictions of game theory (Eq.~\ref{c3}).


Few observations about Eq.~(\ref{eq:defhf}):
\begin{itemize}
  \item it assumes that individuals care only about those they directly interact with, i.e. $p_{i,j}=0$ for all $j\not\in\partial_i$. Also $p_{i,i}=0$. 
  \item It describes a situation in which levels of happiness of individuals may be observable by those they interact with, whereas payoffs may not necessarily be observable. This is consistent with individual behavior being driven by an emotional state -- the happiness -- that internalizes the emotional state of others~\cite{decety2008emotion,singer2006empathic}. With $p_{i,j}>0$ individual $i$ derives satisfaction from taking choices that make $j$ happier, while $p_{i,j}<0$ corresponds to spiteful preferences. 
\item Game theory predictions are invariant under independent affine transformations to each player's payoffs. As for any model of other-regarding preferences, this invariance is lost in Eq.~(\ref{eq:defhf}), which assumes that interpersonal comparison of happiness is possible. Indeed, the value of the baseline status $h_i$ will play an important role in determining the attitude of individuals towards their neighbors, and hence their behavior. 
\end{itemize}

At given levels of pro-sociality encoded in the matrix $\hat P$ with elements $p_{i,j}$, we shall focus on the strategy choices 
\begin{equation}
\label{eq:sistar}
s_i^*(\hat P)\in{\arg}\max_{s_i\in\mathcal{S}_i} u_i(s_i,s_{-i}^*(\hat P);\vec p_i)\,,\qquad\forall i\in\mathcal{N}
\end{equation}
that maximize the happiness of each individual, while others maximize theirs. This is the natural extension of the concept of Nash equilibria (NE). We shall call the profile of choices in Eq.~(\ref{eq:sistar}) simply an equilibrium. We shall restrict attention to equilibria in pure strategies, considering contexts where they exist. 

Levels of pro-sociality are also adjusted in order to maximise individual happiness. In other words, we shall assume that $i$ will care more about $j$ if that makes $i$ happier. 
This can be implemented, e.g., in a dynamics where changes in $p_{i,j}$ are proportional to the gradients
\begin{equation}
\label{eq:duidpij}
\delta p_{i,j}\propto \left.\frac{\partial u_i(s_i,s_{-i})}{\partial p_{i,j}}\right|_{s=s^*}  =-\frac{\partial q_i}{\partial p_{i,j}}
\pi_i(s_i^*,s_{-i}^*)+u_j(s_j^*,s_{-j}^*)+\sum_{\ell\in\partial_i}p_{i,\ell}\frac{\partial u_\ell(s_\ell^*,s_{-\ell}^*)}{\partial p_{i,j}}\,,
\end{equation}
which captures the idea that while behavior ($s_i^*$) may change abruptly, attitudes towards others ($\vec p_i$) change more gradually.
Therefore, the equilibria we shall consider are the simultaneous maxima of $u_i(s_i^*,s_{-i}^*)$ over $\vec p_i$, for all $i\in\mathcal{N}$.

 The last term of Eq.~(\ref{eq:duidpij}) implies that $i$ will care more about $j$ if that makes those ($\ell$) she cares about ($p_{i,\ell}>0$) happier, or those that $i$ dislikes ($p_{i,\ell}<0$) worse off. Our setup is defined by Eqs.~(\ref{eq:defhf},\ref{eq:sistar}) and (\ref{eq:duidpij}). The remainder of the paper is devoted to studying the consequences of these definitions.

\section{General results}

The linear system in Eq.~(\ref{eq:defhf}) can be solved expressing $u_i$ explicitly in terms of $\pi_j$'s
\begin{equation}
\label{eq:uibij}
u_i(s_i,s_{-i})=\sum_{j\in\mathcal{N}} b_{i,j}\left[(1-q_j)\pi_j(s_j,s_{-j})+h_j\right]\,.
\end{equation}
where $b_{i,j}$ is the $i,j^{\rm th}$ element of the matrix 
 \begin{equation}
\hat B=\frac{1}{1-\hat P}=1+\hat P+\hat P^2+\ldots+\hat P^n+\ldots
\end{equation}
In Eq.~(\ref{eq:uibij}), $b_{i,j}$ aggregates the cumulative effects of the propagation of other-regarding attitudes from $i$ to $j$, through neighbors. 
This shows that happiness $u_i$ is, in principle, a collective property that may extend to the whole social network.

This setting entails some general consequences:
\begin{proposition}
    Attitudes of individuals towards neighbors are determined by relative comparisons of one's payoff with others' happiness.
\end{proposition}
This is because, as shown in the Appendix, for all $i\in\mathcal{N}$
\begin{equation}
\label{eq:duidpij1}
\frac{\partial u_i}{\partial p_{i,j}}=b_{i,i}\left[u_j-
\frac{\partial q_i}{\partial p_{i,j}}
\pi_i\right]\,.
\end{equation}
The factor $b_{i,i}=1+\sum_j p_{i,j}p_{j,i}+\ldots$ in Eq.~(\ref{eq:duidpij1}) is a generalized measure of reciprocity, because it accounts for all loops through which the effects of $i$'s concern for others propagate back to her, thereby affecting her happiness. 
Therefore, some effects of reciprocity emerge endogenously in this framework without the need of being introduced exogenously. 

\begin{proposition}
    Spiteful behavior -- $p_{i,j}<0$ for some $i\in\mathcal{N}$ and $j\in\partial_i$ -- can only occur if $h_j<0$ for some $j\in\mathcal{N}$.
\end{proposition}
\begin{proof}
Since $b_{i,i}>0$ (see Appendix), the proof relies on showing that the terms in square brackets of Eq.~(\ref{eq:duidpij1}) are positive when $p_{i,j}<0$. For all $i\in\mathcal{N}$, $\pi_i$ is positive by definition and $\frac{\partial q_i}{\partial p_{i,j}}<0$ for $p_{i,j}<0$ because of Eq.~(\ref{c4}). If $h_j\ge 0$, $u_j$ is also positive, because $j$ can secure a level of $u_j$ at least equal to $\pi_j+h_j\ge 0$ setting $\vec p_j=0$. Therefore, ${ u_i}$ increases with $p_{i,j}$ when $p_{i,j}<0$, so no equilibrium with $p_{i,j}\le 0$ can exist. 
\end{proof}
\noindent
As a corollary, spiteful behavior ($p_{i,j}<0$) in the present setting can only be directed to individuals with $h_j<-\pi_j$.

\begin{proposition}
    In equilibrium, rich individuals tend to be selfish. 
\end{proposition}
\begin{proof}
This is also a consequence of Eq.~\eqref{eq:duidpij1}. In equilibrium, a selfish individual with $\vec p_i=0$ should have $\frac{\partial u_i}{\partial p_{i,j}}$ of the opposite sign of $p_{i,j}$, and hence of $\frac{\partial q_i}{\partial p_{i,j}}$ (by Eq.~\ref{c4}), for all $j\in\partial_i$. This condition holds if $\pi_i>u_j/\frac{\partial q_i}{\partial p_{i,j}}$ for all $j\in\partial_i$, i.e. if $i$ is rich enough.
\end{proof}

A further result can be shown in the case where caring for others doesn't become more costly the more one does it. In this case:

\begin{proposition}
If caring for others (either positively or negatively) becomes less costly the more one does it, there is no interior equilibrium with $0<|p_{i,j}|<1$ for some $i\in\mathcal{N}$ and $j\in\partial_i$.
\label{prop:convexity}
\end{proposition}
\begin{proof}
Taking a further derivative of Eq.~\eqref{eq:duidpij1} one finds
\begin{equation}
\label{eq:d2uidpij2}
\frac{\partial^2 u_i}{\partial p_{i,j}^2}=2b_{j,i}\frac{\partial u_i}{\partial p_{i,j}}-b_{i,i}\pi_i\frac{\partial^2 q_i}{\partial p_{i,j}^2}
\end{equation}
If there is an interior equilibrium $\vec{p}_i^*$ with $0<|p_{i,j}^*|<1$ for some $j\in\partial_i$, the first term of Eq.~\eqref{eq:d2uidpij2} vanishes. Then, non-increasing marginal costs of pro-sociality (i.e. 
$\frac{\partial^2 q_i}{\partial p_{i,j}^2}\le 0$
) imply that $\left.\frac{\partial^2 u_i}{\partial p_{i,j}^2}\right|_{\vec{p}^*}\ge 0$, because both $b_{i,i}$ and $\pi_i$ are positive. Then $\vec{p}_i^*$ cannot be a maximum of $u_i$.
\end{proof}

\begin{proposition}
The Nash equilibrium $\mathbf{s}^*$ is an equilibrium if
\begin{equation}
    \pi_i(s_i^*,s_{-i}^*)\ge\max_{j\in\partial_i}\max\left\{
    -\frac{\pi_j(s_j^*,s_{-j}^*)+h_j}{\chi_{i,j}^-},\frac{\pi_j(s_j^*,s_{-j}^*)+h_j}{\chi_{i,j}^+}\right\}\,,~~\forall i\in\mathcal{N}\,.
\end{equation}
where 
\[
\chi_{i,j}^\pm=\pm\lim_{p_{i,j}\to 0^{\pm}}\left.\frac{\partial q_i}{\partial p_{i,j}}\right|_{p_{i,-j}=0}
\]
and $p_{i,-j}$ stands for all the coefficients $p_{i,k}$ with $k\ne j$. 
If $\frac{\partial^2 q_i}{\partial p_{i,j}^2}\le 0$, these conditions are also sufficient.
\end{proposition}
\noindent 
The proof follows directly from analyzing the stability conditions of the $\hat P=0$ state. This condition restricts the stability of Nash equilibria to the region where $h_i$ are not too large or too small. Close inspection of the stability conditions reveals that prosocial behavior is triggered by large positive values of the status $h_j$ of some of their neighbors, whereas large negative values of $h_j$ promote spiteful behavior among the neighbors of $j$.

As an application of the last two propositions, we note that for $q_i(\vec{p}_i)=\sum_{j\in\partial_i}|p_{i,j}|^\alpha$ with $\alpha<1$ the Nash equilibrium $\vec{p}_i=0$ for all $i\in\mathcal{N}$ is always stable because $\chi_{i,j}^\pm=0$, and there cannot be any interior equilibrium.

\section{Illustrative cases}

Eqs.~(\ref{eq:defhf}, \ref{eq:sistar}) and (\ref{eq:duidpij}) generate a complex phenomenology that we shall explore discussing some illustrative cases. We will focus on the case where 
\begin{equation}
    q_i(\vec p_i)=\frac{1}{\bar p}\sum_{j\in\partial_i}|p_{i,j}|\,,\qquad \bar p\in [0,1]
    \label{eq:qi}
\end{equation}
with $\vec p_i$ constrained to $\sum_{j\in\partial_i}|p_{i,j}|\le \bar p$. This choice condenses the tradeoff between the benefits of other-regarding behavior and payoffs into a single parameter $\bar p$ -- whose inverse can be interpreted as the price of other-regarding behavior. 
This choice also simplifies the analysis because, in view of Proposition~\ref{prop:convexity}, we need only to investigate solutions where each individual $i$ is egoist ($\vec p_i=0,~ q_i=0$), prosocial ($\vec p_i\ge 0,~ q_i=1$) or spiteful ($\vec p_i\le 0,~ q_i=1$). Consequently, we shall partition the set of individuals into three sets $\mathcal{N}=\mathcal{E}\cup\mathcal{P}\cup\mathcal{S}$. This makes the results that we shall present below more easily interpretable. 
 
\subsection{Two players games}

\begin{figure}[h!]
\centering
    \includegraphics[width=.85\textwidth]{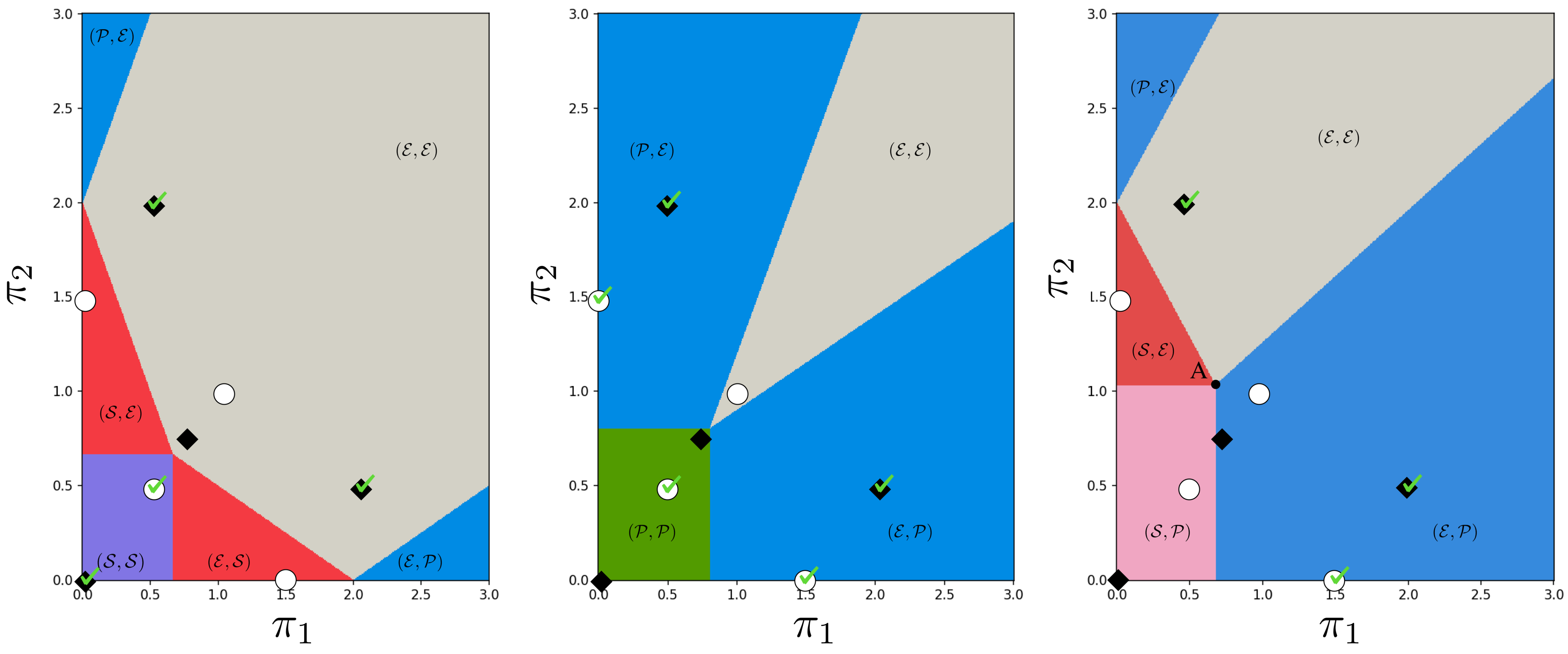}
    \caption{Regions with distinct preferences in $2\times 2$ games. Left $h_1=h_2=-2$, $\bar p=0.5$, center $h_1=h_2=0.8$, $\bar p=0.5$, right $h_1=0.8,~h_2=-2$ and $\bar p=0.7$. Payoff pairs of the prisoners' dilemma are denoted as $\circ$ and those of a hawk-dove game as $\blackdiamond$. Equilibria are marked with a green $\checkmark$.
    }
    \label{fig2x2}
\end{figure}

Much of the intuition on strategic interaction can be gained by studying simple $2\times 2$ games, which are contexts where two players interact and each can chose one of two strategies. In order to see how {\it homo-felix} behaves in these contexts, we shall first partition the positive quadrant spanned by the payoffs $(\pi_1,\pi_2)$ into regions with distinct preferences, as done in Fig.~\ref{fig2x2}. This is obtained analyzing the stability conditions, i.e. 
\begin{eqnarray}
\label{s1}
    \left.\frac{\partial u_i}{\partial p_i}\right|_{p_i=+\bar p}\ge 0~~~ & \Longrightarrow & ~~~i\in\mathcal{P}\\
\label{s2}
    \left.\frac{\partial u_i}{\partial p_i}\right|_{p_i=0^+} \le 0 ~~~&\hbox{and}&~~~\left.\frac{\partial u_i}{\partial p_i}\right|_{p_i=0^-} \ge 0~~~ \Longrightarrow  ~~~i\in\mathcal{E}\\
\label{s3}
    \left.\frac{\partial u_i}{\partial p_i}\right|_{p_i=-\bar p}\le 0~~~ & \Longrightarrow & ~~~i\in\mathcal{S}
\end{eqnarray}
as discussed in the Appendix. 
As the values of $h_1,h_2$ and $\bar p$ vary, the boundaries of the regions change and so different outcomes arise. For example,  
the rectangular regions close to the origin shrink as $h_1$ or $h_2$ decrease and disappear when $h_1, h_2\to 0$, and the $(\mathcal{E},\mathcal{E})$ region shrinks as $(h_1,h_2)$ moves further and further away from the origin. 

The analysis of the equilibria proceeds by adding to this graph the points corresponding to the attainable payoff pairs $(\pi_1,\pi_2)$ for the strategy profiles $\mathbf{s}$ of a specific game. Fig.~\ref{fig2x2} shows these construction for the two games

\begin{align}
\nonumber
\begin{array}{cc} \hbox{Prisoner's Dilemma}& \qquad\qquad\hbox{Hawk Dove} \\
    \left.\begin{array}{c|c|c|} & s_2=0 & s_2=1 \\\hline s_1=0 & 1/2,1/2 & 3/2,0 \\\hline s_1=1 & 0,3/2 & 1,1\\\hline\end{array}\right. &     
    \qquad\qquad\left.\begin{array}{c|c|c|} & s_2=0 & s_2=1 \\\hline s_1=0 & 0,0 & 2,1/2 \\\hline s_1=1 & 1/2,2 & 3/4 ,3/4\\\hline\end{array}\right.
\end{array}
\end{align}

If the strategies $\mathbf{s}$ that correspond to a point $(\pi_1,\pi_2)$ coincide with the solution of Eq.~(\ref{eq:sistar}), i.e. $\mathbf{s}=\mathbf{s}^*(\hat P)$, then this point is an equilibrium otherwise it is not. 
In general, the equilibria are found as fixed points of the best response correspondence for a modified payoff table $\tilde \pi$ that accounts for other-regarding preferences\footnote{The dependence on payoffs vanish when $p_i\to\bar p$ for both players, because $q_i\to 1$. The last two expressions for $\tilde\pi_i$ are obtained setting $q_i=q_{-i}=1-\epsilon$ and then taking the limit $\epsilon\to 0$.}:
\begin{equation}
    b_i(s_{-i})={\rm arg}\max_{s_i\in\mathcal{S}_i}\tilde\pi_i(s_i,s_{-i})\,,~~ \tilde\pi_i(s_i,s_{-i})=\left\{\begin{array}{cl}
        \pi_i(s_i,s_{-i}) & \hbox{if $i\in\mathcal{E}$} \\
        \pi_{-i}(s_{-i},s_{i}) & \hbox{if $i\in\mathcal{P}$ and $-i\in\mathcal{E}$} \\
        -\pi_{-i}(s_{-i},s_{i}) & \hbox{if $i\in\mathcal{S}$ and $-i\in\mathcal{E}$} \\
        \pi_i(s_i,s_{-i})+\bar p\pi_{-i}(s_{-i},s_i) & \hbox{if $i\in\mathcal{P}$ and $-i\not\in\mathcal{E}$}\\
        \pi_i(s_i,s_{-i})-\bar p\pi_{-i}(s_{-i},s_i) & \hbox{if $i\in\mathcal{S}$ and $-i\not\in\mathcal{E}$}
    \end{array}\right.
    \label{eq:bestresponse}
\end{equation}

For example, if the Nash equilibrium falls in the light gray region $(\mathcal{E},\mathcal{E})$ of Fig.~(\ref{fig2x2}) then it is also an equilibrium. 
If a point $(\pi_1,\pi_2)$ falls in the region in which player $1$ is prosocial and player $2$ is not egoist, then $s_1$ must be a maximum of $\pi_1(s_1,s_2)+\bar p\pi_2(s_2,s_1)$ in order for that point to be an equilibrium. 

For the prisoner's dilemma, when $h_i>0$ the Nash equilibrium may not be an equilibrium, and the exploitation of one player by the other is possible if the former has prosocial preferences. When $h_1>0$ and $h_2<0$ the better off player may develop spiteful preferences and exploit the worse off (prosocial) one in equilibrium. For the Hawk-Dove game, the mutual escalation outcome $s_1=s_2=0$ can be an equilibrium when both $h_i$ are negative and players develop spiteful preferences.

The analysis of the equilibria can also proceed by first finding the Nash equilibria of the modified game $\tilde\pi$ for given $(p_1,p_2)$, and then determine what values of $(h_1,h_2)$ would confer stability to them. Let us illustrate this for the ultimatum game. Here player $1$ decides how to split a premium with the other player. She can either offer a minimal amount $0^+$ and keep the rest, or offer a share $a$ of the premium and keep the rest. Player $2$, in turn, can either accept the split or reject it, in which case both players receive nothing. In normal form, this is expressed by the payoff table on the left 
\[
\begin{array}{cc} \qquad\pi(\mathbf{s}) & \qquad\qquad\tilde\pi(\mathbf{s})~~~(1,2\not\in\mathcal{E}) 
\\
    \left.\begin{array}{c|c|c|} & s_2\!=\!0 & s_2\!=\!1 \\\hline s_1\!=\!0 & 0,0 & 1,0^+ \\\hline s_1=1 & 0,0 & 1-a,a\\\hline\end{array}\right. &     
    \qquad\left.\begin{array}{c|c|c|} & s_2\!=\!0 & s_2\!=\!1 \\\hline s_1\!=\!0 & 0,0 & 1,p_2 \\\hline s_1\!=\!1 & 0,0 & 1\!-\!a\!+\!p_1 a,a\!+\!p_2(1\!-\!a)\\\hline\end{array}\right.
\end{array}
\]
The Nash equilibrium is $\mathbf{s}^*=(0,1)$, i.e. player $1$ should offer $0^+$ and player $2$ should accept. 
Individuals with other-regarding preferences will consider instead the modified game $\tilde\pi$, that when $1,2\not\in\mathcal{E}$ is the one above on the right. 
When $p_2<0$ the best response of player $2$ to $s_1=0$ is $s_2=0$, i.e. reject. Therefore the equilibrium for $p_2<0$ is $\mathbf{s}^*=(0,0)$. 
The stability conditions (\ref{s1}, \ref{s2}) and (\ref{s3}) show that there are two equilibria leading to this outcome: either the $(\mathcal{P},\mathcal{S})$ equilibrium ($p_1=\bar p$ and $p_2=-\bar p$), which is stable as long as $\bar p h_1\le h_2\le-h_1/\bar p$, or the $(\mathcal{S},\mathcal{S})$ equilibrium ($p_1=p_2=-\bar p$), which is stable if $h_2\le\max\{\bar p h_1,h_1/\bar p\}$. The $(\mathcal{P},\mathcal{S})$ equilibrium requires $h_1\le 0$ and that the status of the proposer is lower than that of the accepter ($h_1\le h_2$). In the $(\mathcal{S},\mathcal{S})$ equilibrium the status of $1$ and $2$ is reversed ($h_1>h_2$).

\begin{figure}[h!]
\centering
    \includegraphics[width=.85\textwidth]{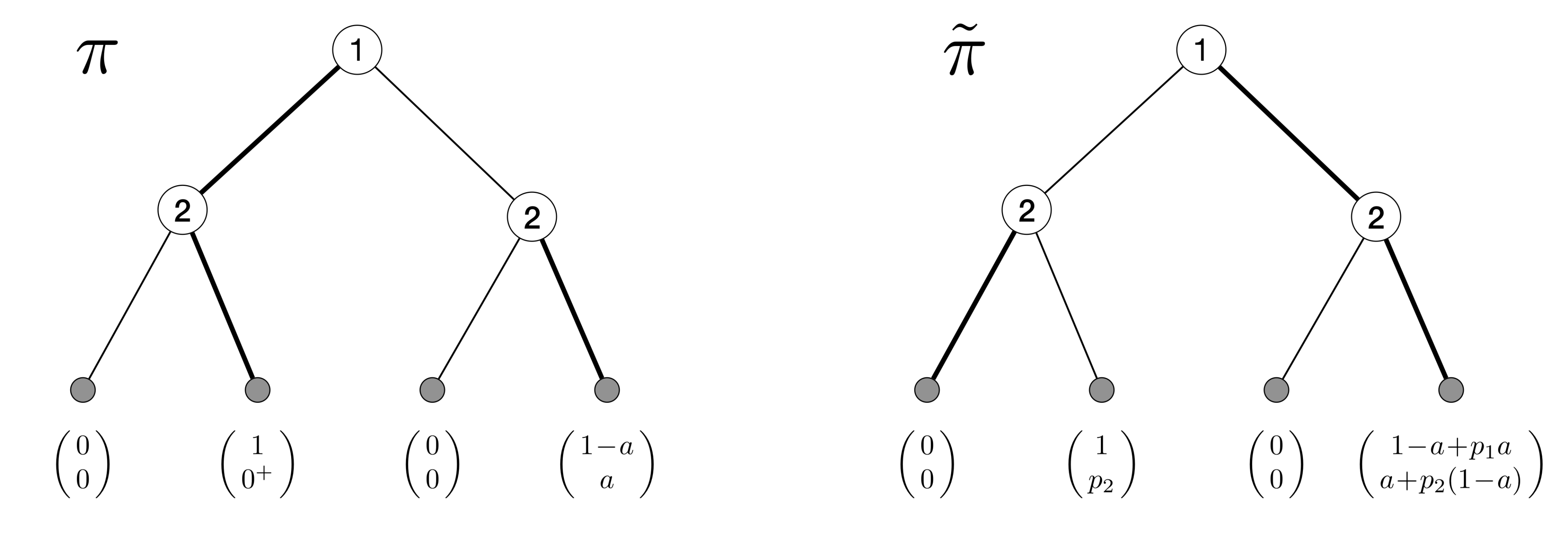}
    \caption{Ultimatum game in extended form (left) and modified game with other-regarding preferences (right). Players move to the left if $s_i=0$ and to the right if $s_i=1$. Best responses are shown as thick lines.
    }
    \label{fig:ext_ug}
\end{figure}

It is worth noticing that these conclusions apply to the game in normal form, i.e. assuming that the game is played simultaneously by the two players. This describes situations where $2$ has to decide whether to accept or reject without knowing the offer. When the accepter knows what the proposer offered before deciding, the game is best described in extensive form, as in Fig.~\ref{fig:ext_ug}. Under the modified game, if player $1$ knows that she's facing a spiteful opponent ($p_2<0$), she can anticipate that $2$ will reject the offer of $0^+$. It will then be convenient for her to offer $a$, in order to secure a positive payoff. Player $2$, in turn, will accept the offer, as long as $a\ge \frac{\bar p}{1+\bar p}$, because otherwise $\tilde\pi_2(s_1\!=\!1,s_2\!=\!1)<0$ and $2$'s best response would again be to reject.

The stability conditions dictate that the outcome $\mathbf{s}=(1,1)$ will arise as a $(\mathcal{P},\mathcal{S})$ equilibrium if $\frac{1+\bar p^2}{\bar p}(1-a)+\bar p h_1\le h_2\le -\frac{1+\bar p^2}{\bar p^2}a- \frac{h_1}{\bar p}$ or as an $(\mathcal{S},\mathcal{S})$ equilibrium if $\frac{1-\bar p^2}{\bar p^2}a+ \frac{h_1}{\bar p}\le h_2\le -\frac{1-\bar p^2}{\bar p}(1-a)+\bar p h_1$. Both cases require that $h_1\le -1-\frac{1-\bar p}{\bar p}a$.

\subsection{$n$ players games}

Finding the equilibria for $n$ person games becomes more complicated. Therefore we shall analyze few simple cases that illustrate the general results discussed in the previous Section. When $q_i(\vec p_i)$ is given by Eq.~(\ref{eq:qi}) the equilibria are attained at corner solutions. This means that for heterogeneous individuals, i.e. when $h_i$ are all different, each individual's attitude in equilibrium is either egoistic towards all her peers ($\vec p_i=0$) or it is prosocial or spiteful towards only one individual, i.e. $p_{i,j}=0$ for all $j\ne i_0(i)$ and $p_{i,i_0(i)}=\pm\bar p$. Hence
\begin{equation}
    u_i(s_i^*,s_{-i}^*)=\max\left[\pi_i(s_i^*,s_{-i}^*),\bar p\max_{j\in\partial_i} u_j(s_j^*,s_{-j}^*),-\bar p\min_{j\in\partial_i} u_j(s_j^*,s_{-j}^*)\right]+h_i\,.
\end{equation}
The first argument in the square brackets above attains when $i\in\mathcal{E}$ ($\vec p_i=0$), the second when $i\in\mathcal{P}$, and the third when $i\in\mathcal{S}$. 
The recipient $i_0(i)$ of other-regarding behavior is the solution of the maximization $i_0(i)={\rm arg}\max_{j\in\partial_i}u_j$ when $i\in\mathcal{P}$, and $i_0(i)={\rm arg}\min_{j\in\partial_i}u_j$ when $i\in\mathcal{S}$. The behavior $s_i^*$ is again determined as the fixed point of the best responses, Eq.~(\ref{eq:bestresponse}), with $-i=i_0(i)$.
\begin{figure}
    \centering
    \includegraphics[width=0.7\linewidth]{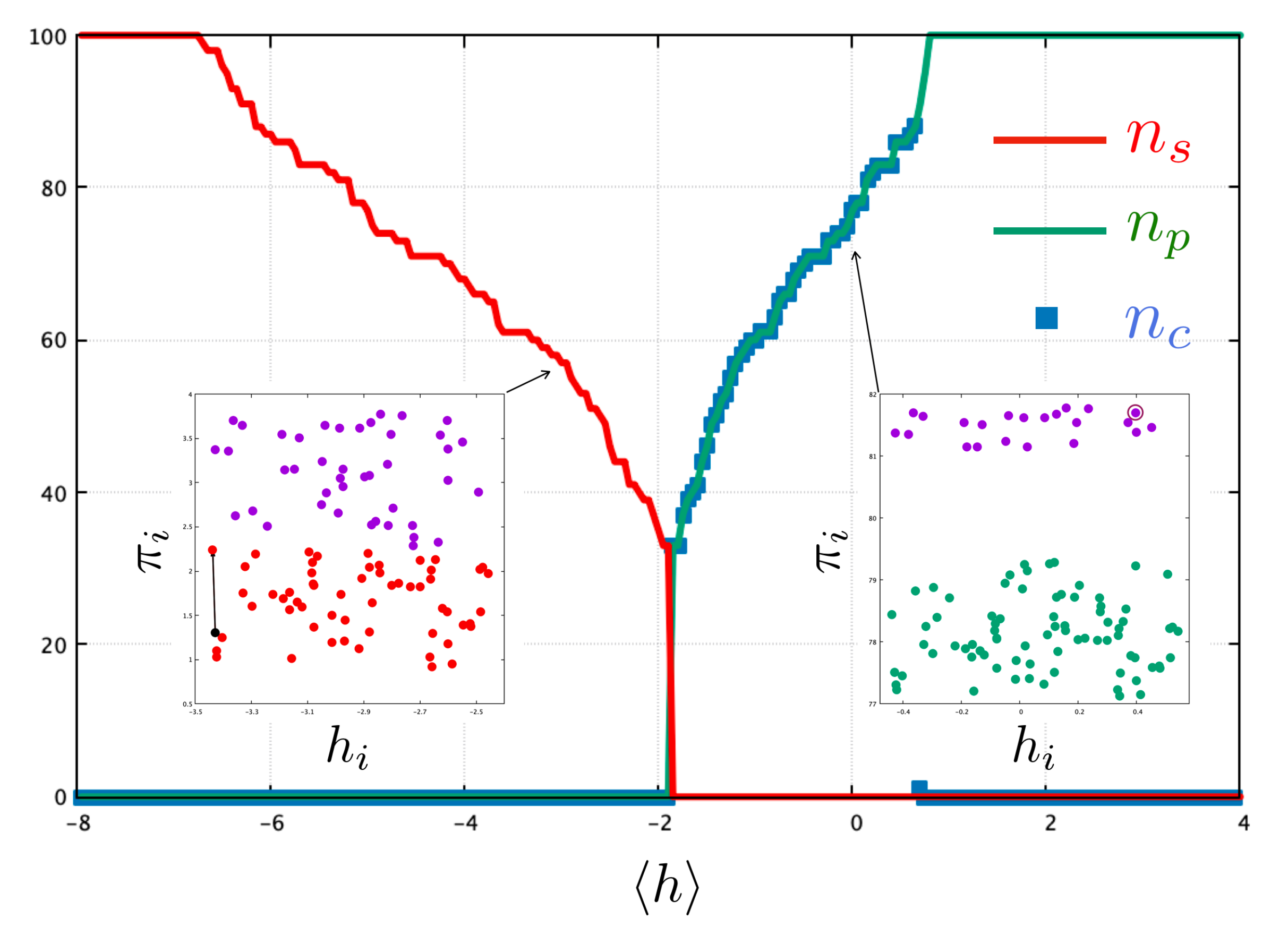}
    \caption{Number of altruists $n_p$ (green line), spiteful $n_s$ (red line) and cooperators $n_c$ (blue points) as a function of the average status $\langle h\rangle$, for a population of $n=100$ agents with $c=0.8$ and $\bar p=0.7$. The parameters $h_i$ are uniformly distributed in the interval $[\langle h\rangle -\frac 1 2,\langle h\rangle +\frac 1 2]$ and the payoffs are given by Eq.~(\ref{eq:piselpd}) with $w_i$ independently distributed uniformly in $[0,3]$. 
    The equilibria are shown for $\langle h\rangle=-3$ (left inset) and $\langle h\rangle=0$ (right inset) in the $(h_i,\pi_i)$ plane. Violet points denote egoists $i\in\mathcal{E}$, green points refer to $i\in\mathcal{P}$ and red ones to $i\in\mathcal{S}$. In the left inset, spiteful behavior is directed towards the individual $i_0$ denoted by a black dot, for all $i\in\mathcal{S}$ but $i_0$, 
    whose spiteful behavior is directed toward the individual marked by an arrow. For the right plot, all $i\in\mathcal{P}$ direct their altruistic behavior towards the circled individual in the right top corner.}
    \label{fig:selective_eq}
\end{figure}
The equilibria are illustrated in Fig.~\ref{fig:selective_eq} for a population of $n=100$ individuals engaged in a common good game ($\partial_i=\mathcal{N}\backslash \{ i \}~~\forall i$) with
\begin{equation}
    \pi_i(s_i,s_{-i})=\sum_{j\ne i}s_j+(1-s_i)c+w_i\,,\qquad s_i=0,1~~~\forall i\in\mathcal{N}
    \label{eq:piselpd}
\end{equation}
as a function of the average status $\langle h\rangle$ (see caption). Eq.~(\ref{eq:piselpd}) describes a situation where individuals can exert an effort ($s_i=1$) that benefits others but not themselves, while paying a cost $c$. $w_i$ is a baseline payoff.

In line with the general results, we find that spite only occurs for negative $\langle h\rangle$, where the behavior conforms to the Nash equilibrium $\mathbf{s}^*=0$. Prosocial and cooperative behavior sets in already for negative $\langle h\rangle$ but, interestingly, cooperation collapses for high values of $\langle h\rangle$ when all $i\in\mathcal{P}$. This occurs when the cost $c$ of cooperation exceeds the marginal benefit $\bar p$ that arises from caring for others.

\subsubsection{Generalized Social Preferences}

Non-egoistic behavior may arise due to a variety of driving factors~\cite{benabou2006incentives}. The cases discussed above relate to the case of {\it selective social preferences}, where other-regarding attitudes $p_{i,j}$ may vary depending on their recipient $j\in\partial_i$.

Social preferences can also encode traits that depend on individual social attributes, such as trustworthiness or reputation, and hence affect in the same way the interaction with all those an individual interacts with. Such generalized social preferences are captured by the choice 
\begin{equation}
p_{i,j}=\left\{\begin{array}{cc}\frac{p_i}{k_i} & j\in\partial_i \\0 & \hbox{else}\end{array}\right.\,,\qquad |p_i|\le\bar p\,,\qquad q_i=\frac{| p_i|}{\bar p}
\end{equation}
where $k_i=|\partial_i|$ is the number of neighbors $i$ interacts with. The analog of Eq.~(\ref{eq:duidpij1}) is easily derived by observing that a change $dp_i$ entails a change $dp_{i,j}=dp_i/k_i$ in all $j\in\partial_i$. Therefore
\begin{equation}
\label{eq:duidqi_gsp}
\frac{\partial u_i}{\partial p_{i}}=\frac{1}{k_i}\sum_{j\in\partial_i}\left.
\frac{\partial u_i}{\partial p_{i,j}}\right|_{p_{i,j}=\frac{q_i}{k_i}}
=b_{i,i}\left(\frac{1}{k_i}\sum_{j\in\partial_i}u_j-\frac{\partial q_i}{\partial p_i}\pi_i\right)\,.
\end{equation}
This implies that a prosocial individual $i$ shall increase $p_i$ if her neighbors' average happiness exceeds a threshold $\pi_i/\bar p$. Likewise anti-social behavior shall be enhanced if the average happiness of neighbors is smaller than $-\pi_i/\bar p$.

\begin{figure}[h!]
\centering
    \includegraphics[width=.75\textwidth, angle=0]{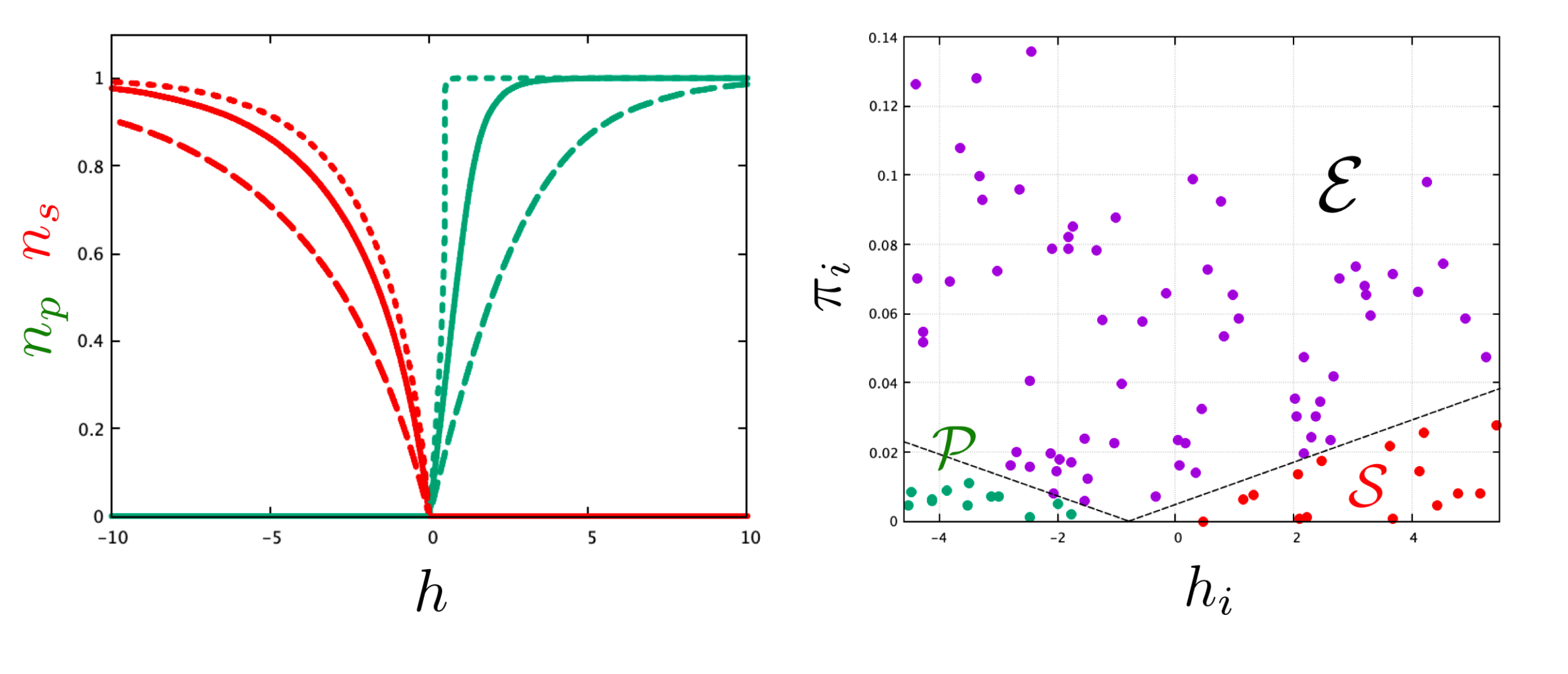}
    \caption{Equilibria of a population with constant $\pi_i$. Left: fraction of prosocial individuals ($n_p$, green) and of spiteful ones ($n_s$, red) in a society ($n=87511$) where $h_i=h$ for all $i$, as a function of $h$, for $\bar p =0.3$ (dashed line), $\bar p = 0.6$ (full line) and $\bar p=0.9$ (dotted line). The values of $\pi_i$ are drawn independently at random from an exponential distribution with unit mean. Either prosocial or spiteful preferences can arise when $h_i=h$ is the same for all $i\in\mathcal{N}$. Right: when $h_i$ is randomly distributed both prosocial and spiteful preferences can arise, for intermediate values of the mean of $h_i$. In the population shown in the right panel, individuals are characterized by the pair $(h_i,\pi_i)$. Those who fall below the dashed line on the left are prosocial ($p_i=\bar p$, green dots), those who fall below the dashed line on the right are spiteful ($p_i=-\bar p$ red dots). All the others are egoists ($p_i=0$ violet dots). The way in which individuals are distributed determines the two boundaries (dashed lines) separating the $\mathcal{P}$, the $\mathcal{S}$ and  the $\mathcal{E}$ regions. The plot was obtained for a society of $n=100$ individuals with $h_i$ uniformly distributed in the interval $[-5.065,4.935]$ and $\pi_i$ exponentially distributed with mean $0.05$ ($\bar p=0.6$). 
    }
    \label{eqdiag}
\end{figure}

In order to illustrate the prevailing equilibria, let us start from the simplest case where $\pi_i$ is  independent of $\mathbf{s}$. We refer to the Appendix for a detailed treatment and focus on the results. In a society where $h_i=h$ attains the same value for all individuals, spiteful behavior arises when $h$ is negative, while when $h$ is positive, prosocial attitudes dominate (see Fig.~\ref{eqdiag} left). In this society, either prosocial or spiteful behaviors can occur, but not both. 

In the presence of heterogeneity, e.g. when both $\pi_i$ and $h_i$ are drawn at random, a similar behavior takes place, as a function of the average value $\langle h\rangle$ of $h_i$. For intermediate values of $\langle h\rangle$ all three types of preferences can be observed, as shown in the right panel of Fig.~\ref{eqdiag}. In this society, the attitudes of individuals are determined by their payoff $\pi_i$ and status $h_i$. Prosocial tendencies are prevalent in worse off individuals, those with low values of $\pi_i$ and $h_i$. Spiteful behaviour characterizes poor (small $\pi_i$) individuals who enjoy high baseline happiness (large $h_i$). The threshold $\bar H$ that divides the two regions is determined self-consistently by
\begin{equation}
    \bar H=\frac{(1-a^2)(\Pi_{\mathcal{E}}+H_{\mathcal{E}})+(1-a)H_{\mathcal{P}}+(1+a)H_{\mathcal{S}}}{1-a^2-a(1-a)n_p+a(1+a)n_s}\,\qquad a=\frac{\bar p}{n-1}
    \label{eq:hbar}
\end{equation}
where $H_{\mathcal{X}}=\sum_{i\in\mathcal{X}}h_i$ for $\mathcal{X}=\mathcal{E},\mathcal{P}$ or $\mathcal{S}$, $\Pi_{\mathcal{E}}=\sum_{i\in\mathcal{E}}\pi_i$, and $n_p$ and $n_s$ are the numbers of altruists and spiteful individuals, respectively. 

Eq.~(\ref{eq:hbar}) shows that the threshold $\bar H$ increases with the general status of the population, therefore when $\bar H>\max_i h_i$ the population counts only prosocial and selfish individuals, while if $\bar H<\min_i h_i$ only spiteful and selfish individuals exist.

Similar conclusions hold in the presence of a strategic component in the dependence of $\pi_i$ on behavior $\mathbf{s}$. 
In particular, a necessary condition for the stability of the Nash equilibrium ($n_p=n_s=0$) is
\begin{equation}
    -\frac{1+\bar p}{\bar p}\Pi_{\mathcal{N}}\le H_{\mathcal{N}}\le \frac{1-\bar p}{\bar p}\Pi_{\mathcal{N}}
    \label{eq:condNE}
\end{equation}
where $\Pi_{\mathcal{N}}=\sum_{j\in\mathcal{N}}\pi_i(s_j^*,s_{-j}^*)$ is the total payoff of the population in the Nash equilibrium (see Appendix). Eq.~(\ref{eq:condNE}) shows that if the average status of the population is too low or too high the Nash equilibrium cannot be stable. 

The analysis of the best response goes as for $2\times 2$ games (see Eq.~\ref{eq:bestresponse}). The payoff $\tilde\pi_i$ that individuals maximize coincides with $\pi_i$ for all $i\in\mathcal{E}$. As long as $\mathcal{E}\ne\emptyset$, $\tilde\pi_i=\langle\pi\rangle_{-i}$ if $i\in\mathcal{P}$ and $\tilde\pi_i=-\langle\pi\rangle_{-i}$ if $i\in\mathcal{S}$, where $\langle\ldots\rangle_{-i}$ denotes an average over all individuals except $i$. In the context of the common good game of Eq.~(\ref{eq:piselpd}), this means that all prosocial individuals will cooperate ($s_i=1$ for all $i\in\mathcal{P}$) and all the others will not ($s_i=0$ for all $i\in\mathcal{E}\cup\mathcal{S}$), as long as there is at least a single egoist ($\mathcal{E}\ne\emptyset$). When $\mathcal{E}=\emptyset$ the modified payoff $\tilde\pi_i$ has to be derived setting $p_i=\bar p(1-\epsilon)$ and then taking the limit $\epsilon\to 0$, as for $2\times 2$ games. The result does not change for $i\in\mathcal{S}$, because they will again defect ($s_i=0$). 
Specializing to the case $h_i=h$ for all $i\in\mathcal{N}$, we find that fully homogeneous equilibrium states $\mathcal{N}=\mathcal{S}$ or $\mathcal{E}$ are possible with 
\begin{eqnarray}
    \mathcal{S}=\mathcal{N} &\hbox{for}~~h<-\frac{1+\bar p}{\bar p}c
    \label{condallS}\\
    \mathcal{E}=\mathcal{N} &\hbox{for}~~-\frac{1+\bar p}{\bar p}c\le h\le \frac{1+\bar p}{\bar p}c\label{condNash}
\end{eqnarray}

The analysis of the best response reveals that, in the homogeneous state $\mathcal{N}=\mathcal{P}$, prosocial individuals cooperate ($s_i=1$) as long as the cost is not too high, i.e. as long as
\begin{equation}
    c\le \bar c=\frac{(n-1)\bar p}{(n-1)(1-\bar p)+\bar p}\,,
    \label{eq:barc}
\end{equation}
but they defect ($s_i=0$) when $c>\bar c$.
This means that the stability condition of the fully prosocial equilibrium $\mathcal{N}=\mathcal{P}$ reads 
\begin{eqnarray}
    h>\frac{1+\bar p}{\bar p}c & \qquad & \hbox{for}~~c>\bar c \label{condallP}\\
    h> \frac{1-\bar p}{\bar p}(n-1)& \qquad & \hbox{for}~~c<\bar c\,.
\end{eqnarray}
It is interesting that the fully prosocial state is stable in a broader region of values of $h$ when $c$ is large, as long as $\bar p$ is not too close to one\footnote{Specifically, this holds for $\bar p< 1-2/(3+\sqrt{4n-3})$.}.

\begin{figure}[h!]
\centering
    \includegraphics[width=.55\textwidth]{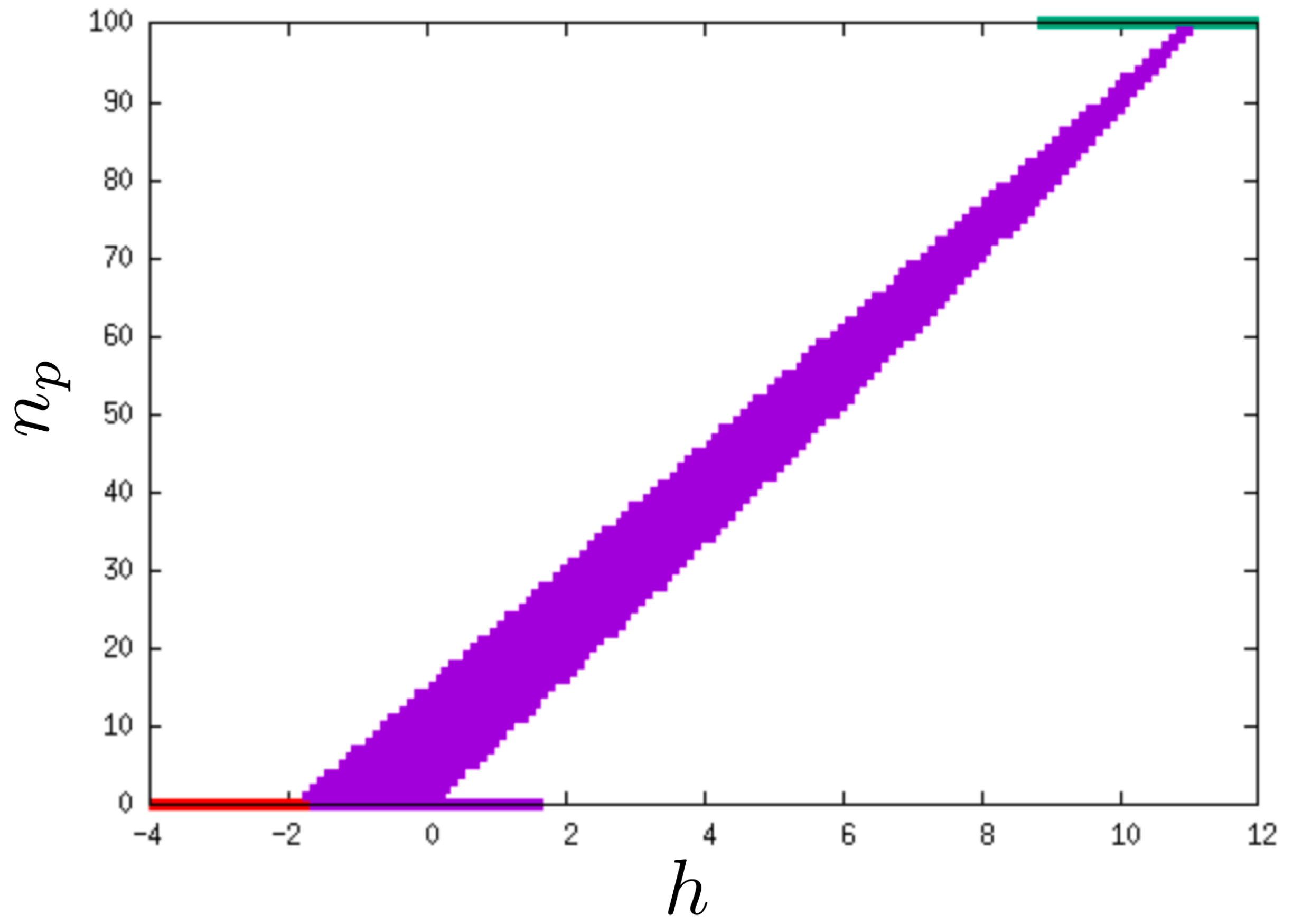}
    \caption{Mixed equilibria for a fully connected society playing a global good game, for $n=100$, $c=0.8$, $\bar p=0.9$ and $h_i=h$ for all $i\in\mathcal{N}$. Each violet point in the $(h,n_p)$ plane marks a possible equilibrium. Red dots correspond to fully spiteful equilibria ($\mathcal{S}=\mathcal{N}$) and green ones to homogeneous prosocial equilibria ($\mathcal{P}=\mathcal{N}$).}
    \label{Fig:mfPD}
\end{figure}

Mixed states with $|\mathcal{P}|=n_p$, $\mathcal{E}=n-n_p$ and $\mathcal{S}=\emptyset$ are also possible in an intermediate region, as shown in Fig.~\ref{Fig:mfPD}.

\section{Conclusions}

The literature on happiness economics has generated a wealth of empirical findings that we believe call for theoretical frameworks that may shed light on the micro-foundations of how social interaction affects subjective well being. The framework discussed in this paper goes in this direction, proposing that subjective well-being is influenced by the subjective well-being of others rather than by their payoffs. This interaction is responsible for feedback loops that generate non-trivial consequences. 

There are several assumptions implicit in this theory. First we assume substitutability between well-being derived from material wealth, social contacts, or status. 
We also do not discuss situations where individuals may decide whom to interact with, as in the endogenous network formation games~\cite{jackson2003strategic}. For example, individuals may seek to acquire more ``importance'' by connecting to ``important'' people, as in Bardoscia {\it et al.}~\cite{bardoscia2013social}. Finally, our framework is purely static and deterministic. As such, it cannot capture prosocial behavior related to expected future outcomes (see e.g.~\cite{dixit2017social}).
Going beyond these limitations points towards interesting lines of further research.

Here, our focus, as in Refs.~\cite{alger2019,tilman2019localized,alger2023evolutionarily,akccay2009theory}, is on how other-regarding preferences develop endogenously within an individual strategic behavior embedded in a specific interaction structure. 
We believe this theory captures fundamental insights that we believe extend beyond this framework, and that warrant investigation on their own.

Specifically, we find that the attitudes of individuals towards their peers are strongly affected by their status ($h_i$) and by the specific trade-off $q_i(\vec p_i$) between material wealth and social reward. In particular, the ``every man against every man'' condition that Hobbes was observing at his time -- a ``state of Nature'' that prevails ``without a common Power to keep them all in awe'' -- arises under very low living standards (i.e. negative status $h_i$) in our framework. Our results suggest that there is no need for a Leviathan to avert generalized conflict by imposing the rule of law. Raising general well-being ($h_i$) is enough to foster pro-social behavior and cooperation in a population of {\it homo-felices}. Our theory also posits that {\it i)} individuals respond to perceived levels of happiness of others and this results in adjustment of pro-social attitudes in terms of up-ward comparison, {\it ii)} rich individuals tend to be selfish while poor ones tend to develop other-regarding preferences, and that {\it iii)} spite is typically directed downward, i.e., towards individuals with negative status $h_i<0$. 

When comparing these predictions with empirical findings in social science or psychology, it is important to realize that human behavior is substantially more nuanced than the sharp conclusions of any theory, in line with Dr Zhivago's warning: ``If you want to know, life is [\ldots] infinitely beyond your or my obtuse theories about it''. In line with our findings, there is substantial evidence that spiteful behavior is significantly more prevalent in societies with weak governance and low standards, while generalized altruistic behavior is more prevalent in societies with high levels of social welfare and development. Piff {\it et al.}~\cite{piff2010having} provides strong evidence that ``the rich are selfish and the poor are altruistic'', but Korndorfer {\it et al.}~\cite{korndorfer2015large} suggest the opposite. A closer look reveals that altruistic behavior differs vastly in terms of type (relational or philanthropic), drivers (e.g. empathy or reputation) and target (immediate peers or institutions), and these differences correlate with wealth~\cite{korndorfer2015large}.
Empirical evidence suggests that spite or envy predominantly affects low status individuals and is directed upward (to the more successful ones), in contrast to our theory. Yet different psychological drives may revert this conclusion. Individuals with a stronger social dominance orientation are more likely to support the enforcement of group-based hierarchies by high-status groups over low-status groups, thereby sustaining status inequalities and potentially restricting opportunities for upward social mobility~\cite{pratto1994social}.

In view of this complexity, the theoretical results discussed in this paper are necessarily an oversimplification that is only partially supported by data. Still, these results indicate how some of the observed behaviors may be related to its determinants, and it suggests avenues for further empirical research\footnote{For example, empirical measures of subjective well-being~\cite{stone2018understanding} focus on individual’s own life satisfaction and barely take into account the interdependence of well-being across individuals.}. 
This, we believe, offers a compelling motivation to further develop and refine this line of theoretical inquiry.

\section{Acknowledgments}

We thank the participants of the TLQS Workshop on Limits to Collective Agency held at the Abdus Salam ICTP (6-10 May 2024) for very interesting and fruitful discussion. We thank in particular Henrik Olsson for interesting comments and discussions. Funding from the National Institute of Oceanography and Applied Geophysics (OGS), within The Laboratory for Quantitative Sustainability, are gratefully acknowledged.

\bibliographystyle{unsrt}
\bibliography{prosociality}

\appendix



\section{On the proof of Proposition 1}

We shall prove a more general of Eq.~(\ref{eq:duidpij1}), i.e. 
\begin{equation}
\label{eq:dukdpij}
\frac{\partial u_k}{\partial p_{i,j}}=b_{k,i}\left[u_j-\frac{\partial q_i}{\partial p_{i,j}}\pi_i\right]\,.
\end{equation}
for all $i,k\in\mathcal{N}$ and $j\in\partial_i$. Taking the derivative of $u_k$, as given by Eq.~(\ref{eq:uibij}) with $i=k$, w.r.t. $p_{i,j}$ we find
\begin{eqnarray}
    \frac{\partial u_k}{\partial p_{i,j}} &=& \sum_\ell\frac{\partial b_{k,\ell}}{\partial p_{i,j}}\left[(1-q_\ell)\pi_\ell+h_\ell\right]-b_{k,i}\frac{\partial q_i}{\partial p_{i,j}}\pi_i\\
    & = & b_{k,i}\left\{\sum_\ell b_{j,\ell}\left[(1-q_\ell)\pi_\ell+h_\ell\right]-\frac{\partial q_i}{\partial p_{i,j}}\pi_i\right]
\end{eqnarray}
where we used the identity $\frac{\partial b_{\ell,k}}{\partial p_{i,j}}=b_{\ell,i}b_{j,k}$. The first term in square brackets of the last equation coincides with $u_j$, again because of Eq.~(\ref{eq:uibij}). This then leads us to Eq.~(\ref{eq:dukdpij}). 

\section{Proof that $b_{i,i}>0$}

Let $\hat P$ be a real $n \times n$ matrix with elements $p_{i,j}$ that satisfies Eq.~(\ref{c2}).
The $i$-th diagonal element of $\hat B=(\hat I - \hat P)^{-1}$ is given by:
\begin{equation}
    b_{ii} = \frac{\det(\hat A_{i,i})}{\det(\hat A)},
    \label{eq:appbii}
\end{equation}
where $\hat A=\hat I-\hat P$ and $\hat A_{i,i}$ denotes the $(n-1) \times (n-1)$ principal submatrix obtained by deleting the $i$-th row and $i$-th column of $\hat A$. In order to prove that the diagonal element is positive, we will show that both the numerator and the denominator are strictly positive.

First, we analyze the eigenvalues of $\hat A$. Let $\lambda$ be an eigenvalue of $\hat P$. According to the {Gershgorin Circle Theorem}~\cite{horn2012matrix}, every eigenvalue of $\hat P$ lies within at least one Gershgorin disc $D_k$ centered at $p_{k,k}$ with radius $R_k = \sum_{j \neq k} |p_{k,j}|$.

Given that $p_{k,k} = 0$ and $\sum_{j} |p_{k,j}| < 1$, every eigenvalue of $\hat P$ satisfies $|\lambda|<1$. 
Hence the eigenvalues of $\hat A = \hat I - \hat P$, that are given by $\mu = 1 - \lambda$, have all positive real values.
The determinant of $\hat A$ is the product of the eigenvalues. Since $\hat A$ is a real matrix, its eigenvalues occur either as real numbers or as conjugate pairs. The product of conjugate pairs is positive, so it is enough that real eigenvalues of $\hat A$ be positive, which is true, as shown above. 
Therefore, the denominator in Eq.~(\ref{eq:appbii}) is strictly positive, i.e. $\det(\hat A) > 0$. 

The same is true for the numerat or, because $\hat A_{i,i}$ satisfies the same conditions of $\hat A$. Indeed $\hat A_{ii}=\hat I' - \hat P'$, where $\hat I'$ is the identity matrix of size $(n-1)$ and $\hat P'$ is the principal sub-matrix of $\hat P$ obtained by removing row $i$ and column $i$. $\hat P'$ satisfies the same conditions as $\hat P$ (zero diagonal elements and row sums of absolute values less than one). Hence also $\det(\hat A_{i,i})>0$.
Thus, the diagonal elements of $\hat B=(\hat I-\hat P)^{-1}$ are strictly positive.

\section{$2\times 2$ games}

In a population of $n=2$ individuals the vector $\vec p_i$ reduces to just one coefficient $p_i\equiv p_{i,\bar i}$, where $\bar i=2$ if $i=1$ and $\bar i=1$ if $i=2$. The happiness can  be written explicitly in terms of payoffs and of $h_i$ as
\begin{equation}
    u_i=\frac{(1-q_i)\pi_i+h_i+p_i\left[(1-q_{\bar i})+h_{\bar i}\right]}{1-p_1p_2}
\end{equation}
With $q_i=|p_i|/\bar p$
\begin{equation}
    \frac{\partial u_i}{\partial p_i}=b_{i,i}\left[u_{\bar i}-\frac{\pi_i}{\bar p}\right]
\end{equation}
Let us now consider the different equilibria:
\begin{description}
    \item[- $1,2\in\mathcal{P}$]  then $p_1=p_2=\bar p$ and $q_1=q_2=1$. Stability requires that  $\frac{\partial u_i}{\partial p_i}>0$ , i.e. 
    \[
    \pi_i<\frac{\bar p}{1-\bar p^2}\left(\bar p h_i+h_{\bar i}\right)\,,\qquad i=1,2.
    \]
    \item[- $i\in\mathcal{P}$ and $\bar i\in\mathcal{E}$] Then $p_i=\bar p$, $q_i=1$ and $p_{\bar i}=q_{\bar i}=0$. The condition $\left.\frac{\partial u_i}{\partial p_i}\right|_{p_i=\bar p}>0$ implies $\pi_i<\bar p(\pi_{\bar i}+h_{\bar i})$ while for $\bar i$ both the derivatives of $u_i$ w.r.t. $p_{\bar i}$ as $p_{\bar i}\to 0^\pm$ have to be considered. This leads to
    \[
    \pi_{\bar i}>\max\left[\frac{\bar p}{1-\bar p^2}(h_i+\bar ph_{\bar i}),-\frac{\bar p}{1+\bar p^2}(h_i+\bar ph_{\bar i})\right]
    \]
    \item[- Nash equilibrium $1,2\in\mathcal{E}$]. For $p_1=p_2=q_1=q_2=0$ the stability conditions read
    \[
    \pi_i>\bar p|\pi_{\bar i}+h_{\bar i}|\,,\qquad i=1,2
    \]
    \item[- $i\in\mathcal{S}$ and $\bar i\in\mathcal{E}$] Then $p_i=-\bar p$, $q_i=1$ and $p_{\bar i}=q_{\bar i}=0$. The condition $\left.\frac{\partial u_i}{\partial p_i}\right|_{p_i=-\bar p}<0$ implies $\pi_i<-\bar p(\pi_{\bar i}+h_{\bar i})$. The same analysis as above leads to
    \[
    \pi_{\bar i}>\max\left[\frac{\bar p}{1+\bar p^2}(h_i-\bar ph_{\bar i}),-\frac{\bar p}{1-\bar p^2}(h_i+\bar ph_{\bar i})\right]\,.
    \]
    \item[- $1,2\in\mathcal{S}$] $p_1=p_2=\bar p$ and $q_1=q_2=1$. Stability requires
    \[
    \pi_i<-\frac{\bar p}{1+\bar p^2}\left(h_{\bar i}+\bar p h_i\right)\,,\qquad i=1,2\,.
    \]
    \item[- $i\in\mathcal{P}$ and $\bar i\in\mathcal{S}$] $p_i=-p_{\bar i}=\bar p$ and $q_i=q_{\bar i}=0$. Equilibrium requires
    \[
    \pi_i<\frac{\bar p}{1+\bar p^2}(h_{\bar i}-\bar p h_i)\,,\qquad \pi_{\bar i}<-\frac{\bar p}{1+\bar p^2}(h_i+\bar p h_{\bar i})\,.
    \]
\end{description}
Concerning the behavior $s_i^*(\hat P)$ in equilibrium, notice that $u_i$ is independent of $\mathbf{s}$ when $1,2\not\in\mathcal{E}$, because $q_1=q_2=1$. Then we consider the best response when $p_i=\pm(1-\epsilon)\bar p$ and $q_i=1-\epsilon$, in the limit $\epsilon\to 0$. This leads to the last two cases in Eq.~(\ref{eq:bestresponse}).

\section{Fully connected system with generalized attitudes}

In this Section we study the case $p_{i,j}=a_i$ for all $j\neq i$ and $p_{i,i}=0$ ($|a_i|< 1/(n-1)$). Then
\begin{equation}
    u_i=(1-q_i)\pi_i+a_i\sum_{j\ne i}u_j+h_i=\sum_jb_{i,j}[(1-q_j)\pi_j+h_j]\,.
\end{equation}
The inverse of $1-P$ is
\begin{equation}
    b_{i,j}=\frac{a_i}{(1-A)(1+a_i)(1+a_j)}+\frac{\delta_{i,j}}{1+a_i}\,,\qquad A=\sum_j \frac{a_j}{1+a_j}\,,
\end{equation}
so that
\begin{equation}
    u_i=\frac{(1-q_i)\pi_i+h_i}{1+a_i}+\frac{(n-1)a_i}{(1-A)(1+a_i)}\,,
\end{equation}
Also
\begin{equation}
    \frac{\partial u_i}{\partial p_i}=\frac{b_{i,i}}{(n-1)(1+a_i)}\left\{\bar H-h_i-\left[1-q_i+({n-1})(1+a_i)q_i'\right]\pi_i\right\}\,,\qquad \bar H=\sum_{j\in\mathcal{N}}\frac{(1-q_j)\pi_j+h_j}{(1-A)(1+a_j)}
\end{equation}
and $q_i'=\left.\frac{\partial q_i}{\partial p_{i,j}}\right|_{p_{i,j}=(n-1)a_i}$.

We focus on the model $q_i=(n-1)|a_i|/\bar p$ with $a_i\in[-\frac{\bar p}{n-1},\frac{\bar p}{n-1}]$ and $\bar p<1$.

\paragraph{Prosocial individuals} have $a_i=(n-1)\bar p$, $q_i=1$. 
\begin{eqnarray}
    u_i&=&\frac{(n-1)h_i+\bar p\bar H}{n-1+\bar p}\\
    \left.\frac{\partial u_i}{\partial p_i}\right|_{p_i=\bar p} &=& \frac{b_{i,i}}{n-1+\bar p}\left[\bar H-h_i-\frac{n-1+\bar p}{\bar p}\pi_i\right]
    \end{eqnarray}
Therefore, the payoff of a prosocial individual has to satisfy
\begin{equation}
    \pi_i\le \frac{\bar p\left(\bar H-h_i\right)}{n-1+\bar p}\,.
    \label{eq:condPpi}
\end{equation}

\paragraph{Egoists} have $a_i=q_i=0$. The happiness is $u_i = \pi_i+h_i$ and the stability conditions require
\begin{eqnarray}
    \left.\frac{\partial u_i}{\partial p_i}\right|_{p_i=0^+} &=& \frac{1}{n-1}\left[\bar H-h_i-\frac{n-1+\bar p}{\bar p}\pi_i\right]\le 0\\
    \left.\frac{\partial u_i}{\partial p_i}\right|_{p_i=0^+} &=& \frac{1}{n-1}\left[\bar H-h_i+\frac{n-1-\bar p}{\bar p}\pi_i\right]\ge 0
\end{eqnarray}
Therefore, the payoff of a prosocial individual has to satisfy
\begin{equation}
    \pi_i\ge \max\left\{\frac{\bar p\left(\bar H-h_i\right)}{n-1+\bar p},\frac{\bar p\left(h_i-\bar H\right)}{n-1-\bar p}\right\}\,.
\end{equation}

\paragraph{Spiteful ones} have $a_i=-(n-1)\bar p$ and $q_i=1$. The payoff is 
\[
u_i = \frac{(n-1)h_i-\bar p \bar H}{n-1-\bar p}
\]
and the stability conditions require
\[
    \left.\frac{\partial u_i}{\partial p_i}\right|_{p_i=-\bar p} = \frac{b_{i,i}}{n-1-\bar p}\left[\bar H-h_i+\frac{n-1-\bar p}{\bar p}\pi_i\right]\le 0\\
\]
that is

\begin{equation}
    \pi_i\le \frac{\bar p(h_i-\bar H)}{n-1-\bar p}
\label{condspite}
\end{equation}

\paragraph{Equilibria}

If there are $n_p=|\mathcal{P}|$ prosocial and $n_s=|\mathcal{S}|$ spiteful individuals, then 
\[
A=\frac{n_p\bar p}{n-1+\bar p}-\frac{n_s\bar p}{n-1-\bar p}\,,
\]
and $\bar H$ is given by Eq.~(\ref{eq:hbar}) in the main text. The equilibrium is characterized as in Fig.~\ref{eqdiag}.

\paragraph{The Nash equilibrium} ($n_p=n_s=0$) has $a_i=A=0$ and $\bar H=\Pi_{\mathcal{N}}+H_{\mathcal{N}}$, where we used the notation $X_{\mathcal{N}}=\sum_{i\in\mathcal{N}}x_i$. For all $i\in\mathcal{N}$ it must be that
\begin{equation}
    \pi_i\ge \frac{\bar p}{n-1+\bar p}\left(\Pi_{\mathcal{N}}+H_{\mathcal{N}}-h_i\right)~~\hbox{and}~~\pi_i\ge \frac{\bar p}{n-1-\bar p}\left(h_i-\Pi_{\mathcal{N}}-H_{\mathcal{N}}\right)
\end{equation}
Summing over $i\in\mathcal{N}$, we find the necessary condition Eq.~(\ref{eq:condNE}) for the stability of the Nash equilibrium.

\subsection{Common good game}

Let us consider a population playing the game of Eq.~(\ref{eq:piselpd}) with $h_i=h$ and $w_i=0,~\forall i\in\mathcal{N}$. 

In the Nash equilibrium $a_i=s_i=0~~\forall i\in\mathcal{N}$. Therefore $\pi_i=c$ and $\bar H=H_{\mathcal{N}}+\Pi_{\mathcal{N}}=n(h+c)$. The stability condition can be rewritten as 
\[
c\ge\max\left\{\frac{\bar p[nc+(n-1)h]}{n-1+\bar p},-\frac{\bar p[nc+(n-1)h]}{n-1-\bar p}]\right\}
\]
which, after some algebra can be rewritten as Eq.~(\ref{condNash}). 

In the state $\mathcal{S}=\mathcal{N}$ also $s_i=0$ and $\pi_i=c~~\forall i\in\mathcal{N}$, while $a_i=-\bar p/(n-1)$. This is because $\tilde\pi_i$ decreases with $s_i$. Then Eq.~(\ref{condspite}) leads to Eq.~(\ref{condallS}). 

The state $\mathcal{P}$ needs more care because 
\[
\tilde \pi_i(s_i,s_{-i})=\bar \pi+\frac{(n-1)\bar p-[(n-1)(1-\bar p)+\bar p}{(n-1)(1-\bar p)}s_i
\]
where $\bar\pi$ is independent of $s_i$. The best response of prosocial agents then is $s_i=1$ as long as $c<\bar c$ (see Eq.~\ref{eq:barc}).
For $c>\bar c$ the best response is $s_i=0$ for all $i\in\mathcal{N}$ and hence Eq.~(\ref{eq:condPpi}) with $\pi_i=c$ leads to Eq.~(\ref{condallP}). If instead $c<\bar c$, one has to consider Eq.~(\ref{eq:condPpi}) with $\pi_i=n-1$, which leads to the condition $h>(n-1)(\bar p^{-1}-1)$.

\end{document}